\documentclass{article}

\def\ket#1{|{#1}\rangle}

\newcommand{\eq}[1]{Eq.~(\ref{#1})}

\def\mod{\mathop{\mbox{mod}}}

\begin{document}
 
\title{Tools for Quantum Algorithms}
\author{Tad Hogg\\{\tt hogg@parc.xerox.com} \and Carlos Mochon\\{\tt ~~~carlosm@mit.edu~~~}  \and Wolfgang Polak\\{\tt polak@pal.xerox.com} \and Eleanor Rieffel\\{\tt rieffel@pal.xerox.com}}
\maketitle

\begin{abstract}
We present efficient implementations of a number of operations for
quantum computers. These include controlled phase adjustments of the
amplitudes in a superposition, permutations, approximations of transformations
and generalizations of the phase adjustments to block matrix transformations.
These operations generalize those used in proposed quantum search
algorithms.
\end{abstract}
 
\section{Introduction}
\label{introduction}
Shor's factorization algorithm~\cite{Shor-94,Shor-95} and Grover's
search algorithm~\cite{STOC::Grover1996, Grover-xxx} demonstrate that
quantum computers can solve certain problems faster than classical
computers.  It has been well-known for over a decade that any
classical algorithm has a quantum analog of comparable complexity
~\cite{Benioff-80, Benioff-82b, Deutsch-85,Deutsch-89}, and
quantum analogs of classical building blocks have been
studied~\cite{Barenco-et-al-95a, Bernstein-Vazirani-93,
Vedral-et-al-95}.  But to exploit the power of quantum computers and
create algorithms of new complexity classes, we need to use building
blocks that do not have classical analogs but instead take advantage
of quantum parallelism through modifying and mixing amplitudes in
superpositions.

Two sorts of tools have been used effectively in the quantum
algorithms that have been developed so far. First, 
transformations that mix amplitudes, such as the Fourier and Walsh
transforms. Second, selective adjustment of the phases of certain
states that, when combined with a mixing transform, promote amplitude
cancellation or amplification. Such phase adjustments form the basis
of search algorithms for NP problems~\cite{cerf98,Grover-xxx,Hogg-98}.
Here, we discuss efficient implementations of relative phase changes
and of mixing transformations that combine amplitude from only a small
number of states. The choice of phases and which states to mix depends
on a classically efficiently computable function $f$.  As we are
dealing with tools for algorithms in general, specific problems will
not be addressed, so $f$ will remain necessarily abstract.  We discuss
implementations of phase changes, of permutations, of approximations
of transformations, and generalizations of the phase change techniques
to block matrix transformations.  For each of these transformations, we
describe the resources in terms of time, number of calls to $f$, and
number of additional qubits needed for the implementation.
Our aim is simply to describe a collection of efficiently
implementable transformations which we hope will allow future
designers of quantum algorithms to take a somewhat more high level
approach when thinking about how to take advantage of quantum
parallelism. Furthermore, the implementations we describe for more
general operations than have been used in algorithms proposed to date
may form the basis for more effective algorithms.



We assume that the reader is familiar with quantum computing and the 
standard terminology and notation of that field. For an introduction
to the field, see \cite{Rieffel-Polak-98}.


\subsection{General set-up}
\label{setup}

Throughout we will be describing transformations of an $n$ qubit
system.  In order to implement these transformations we will assume at
times that we also have access to an $m$ qubit register in which we
can store values which will help us perform the desired
transformation.  We are particularly interested in describing
transformations that can be efficiently implemented, where by
``efficient'' we mean that the implementation takes a number of steps
that is polynomial in $n$.

We first concern ourselves with transformations that change the
relative phases of components that make up a superposition. Such
transformations correspond to acting on the state with a diagonal
matrix $D$.  Conversely, because quantum operations are unitary, any
operation described by a diagonal matrix will consist of such phase
adjustments.  Since a global phase change has no physical meaning, so
the matrix is only well defined up to multiplication by a constant.
To specify a general phase change would require specifying all $N=
2^n$ elements $D_{xx}$ of the diagonal matrix $D$. Only phase changes
that can be expressed in a concise form are practical.  For this
reason, we will assume that the phase changes are determined by an
efficiently computable function $f$. For example, the function $f(x)$
for Grover's search algorithm computes whether or not $x$ is one of
the desired elements.  In Hogg's algorithms, $f(x)$ depends for
instance on the number of conflicts a state $x$ has with the
constraints and on the size of $x$.  Here, we will take a general $f$
that is efficiently computable classically.

At first glance, the problems we are concerned with may appear
trivial.  How hard could it be to implement a diagonal matrix?
However, these are $2^n\times 2^n$ matrices, and we are interested in
implementing them in a number of steps which is only polynomial in
$n$.  Furthermore, there are many families of transformations that
cannot be efficiently computed, even when they can be described in
terms of an $f$ that can be efficiently implemented. To illustrate
this point we describe a permutation that can be concisely described
in terms of $f$, but which cannot be efficiently implemented.

\label{exchange}
Imagine we are in the simplest set-up for Grover's search algorithm,
where we are looking for a single item in an unstructured database of
size $N=2^n$. The efficiently computable function $f(x)$ simply checks
if $x$ is the desired item, so $f(x)=1$ when $x$ is the desired item
and $f(x)=0$ otherwise. One way to find the item would be to use a
tranformation which switched the state $\ket{00\dots 0}$ with the
state $x$ with $f(x)=1$. If such a transformation could be efficiently
implemented, we could find the desired item much more quickly than
Grover's algorithm does, simply by starting with $\ket{00\dots 0}$,
applying the transformation, and then reading the output, which would
be the desired state. However, as Grover's algorithm is optimal
\cite{Boyer-et-al-96}, this transformation cannot be efficiently
implemented.

Throughout this paper, we use the fact that efficiently implementable
classical functions can be implemented with comparable complexity on a quantum 
computer using standard building blocks.
\cite{Deutsch-85, Bernstein-Vazirani-93, Barenco-et-al-95a}
We assume perfect operations, so we do not deal with error control.
In this paper a
phase change of $e^{\frac{2\pi i}{m}}$ will be treated as one step no
matter how large the $m$.

Let $f(x)$ be a classical polynomially computable function. Quantum
parallelism can be used to compute all the values of $f(x)$
for all $x$ at the same time. 
This computation uses an additional register to 
hold the values of $f$.  We will ignore any temporary workspace which
returns to its original state by the end of the computation that might
be needed to compute $f$.
We use the following standard
transform to implement the quantum parallel computation of $f(x)$, 
\begin{equation}\label{Uf}
U_f:  \ket {x, a} \to \ket {x, a \oplus f(x)},
\end{equation}
where $\oplus$ is the bitwise exclusive-{\sc or}. 
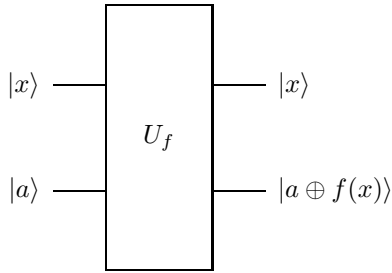
\begin{center}
\begin{picture}(10,10)(0,0)
\put(3,0){\framebox(4,10){$U_f$}}
\put(1,3){\line(1,0){2}}
\put(1,7){\line(1,0){2}}
\put(7,3){\line(1,0){2}}
\put(7,7){\line(1,0){2}}
\put(0.5,7){\makebox(0,0)[r]{$\ket x$}}
\put(0.5,3){\makebox(0,0)[r]{$\ket a$}}
\put(9.5,7){\makebox(0,0)[l]{$\ket x$}}
\put(9.5,3){\makebox(0,0)[l]{$\ket {a\oplus f(x)}$}}
\end{picture}
\end{center}
Consider a superposition of $x$ values,
$$\sum_x a_x \ket{x}.$$ 
Then \eq{Uf} transforms $\sum_x a_x \ket{x} \otimes \ket{0}$ as  
\begin{equation}\label{transform}
	\sum_x a_x \ket{x,0} \rightarrow \sum_x a_x \ket{x, f(x)}.
\end{equation}

\subsection{A summary of the techniques described}


In implementing quantum algorithms it will be useful to have
a variety of techniques depending on whether number of bits or coherence
time (number of operations) is the main limiting factor.

This paper describes several methods for implementing relative phase
changes to components of an $n$-qubit quantum state, which
can be represented as $2^n\times 2^n$ diagonal matrices
$D$.  Specifically if the phase $D_{xx}$ depends on an efficiently computable
function $f(j)$ then
\begin{itemize}
\item if there are only $k$ distinct phase values, $D$ can be implemented in 
$O(k)$ steps and two evaluations of $f$.  The
technique requires ${\lceil \log_2(k) \rceil}$ additional qubits.
\item the well known technique 
for inverting the phase of states selected by $f(x) = 1$
can be extended to change the phase of selected states by a single phase
value which is a $2^m$th root 
of unity.  This extension requires at most $m$ evaluations of $f$, 
an average of less than $2$ evaluations,
and one additional qubit.
\item if all phases in $D$ are multiples of a $k$th root of unity, $D$ can be 
implemented with a single application of $f$ using ${\lceil \log_2(k) \rceil}$
additional qubits and  only $O(\log_2(k))$ operations to prepare these
additional qubits.
\item if the phases in $D$ need only be computed to $k$ bit 
binary precision, $D$ can be implemented in $O(k)$ operations
using one additional qubit and $k$ function evaluations.
\item if $D$ is decomposable, in that it can be written as tensor product of 
single qubit rotations, it can be implemented trivially in $O(n)$ steps
without any additional qubits or function calls.
We give a sufficient and necessary condition for the decomposability of $D$.
\end{itemize}

The utility of diagonal matrices is enhanced if it is possible to perform
permutations on the quantum state efficiently.  We present a technique for
implementing an arbitrary permutation $g$ on a $n$-bit quantum state by one evaluation of $g$ and
one evaluation of $g^{-1}$ using $n$ additional qubits.

Finally, we show how some of the implementation techniques for diagonal
matrices and permutations can be extended to block diagonal matrices, 
which effect amplitude mixing among a small number of states.

\subsection {Related Work}

\label{related-work}
In \cite{Hoyer-97} H{\o}yer shows how to efficiently implement certain unitary
transformations that can be represented as generalized Kronecker products.  
The technique applies to general transformations along the lines of
the quantum Fourier transformation.
His paper includes an efficient implementation for certain permutations and 
and an implementation block diagonal matrices that is similar to 
the one described in section \ref{blocks}.
 
Knill \cite{Knill-95} discusses the approximation of quantum transformations
and proves an
upper bound on the complexity of implementing arbitrary unitary transformations.
The upper bound, while smaller than previous known results, is still exponential
in the number of qubits.  Knill also shows that 
arbitrary unitary transformations cannot be efficiently approximated.  
 
Tucci \cite{Tucci-98} defines a ``quantum compiler'' based on 
Cosine/Sine decomposition of a given unitary matrix.  In principle 
his approach seems promising.  However in its present form the 
quantum compiler takes the actual matrix as input as opposed to
symbolic input, so the space and time complexity just for the input is
exponential in the number of quantum bits $n$.  Furthermore, the current
algorithm rarely generates polynomial implementations even 
when that is possible.

\section{Independent Phase Changes}
\label{independent}
In this section we discuss the efficiency of implementing  phase
changes on components of an $n$-qubit state represented by an
$N\times N$ diagonal matrix $D$ with diagonal entries $d_x$ for
$x$ ranging from $0$ to $N-1$, where $N=2^n$.
The methods vary in their restrictions on the $d_x$'s, their efficiency
in terms of number of operations and calls to $U_f$ needed, and the
number of additional qubits required.

In the worst case a diagonal 
matrix $D$ of size $N \times N$ can be implemented in $O(N)$ steps
by iterating the following procedure over all $N$ values:  For any $x$, 
let $\delta_y(x)$ be the function that is 1 when $y=x$ and 0 otherwise. 
Apply $U_{\delta_y}$ using \eq{Uf} to the original 
state $\sum_x a_x\ket{x,0}$ to
get $\sum_x a_x\ket{x,\delta_y(x)}$.
Then multiply the state by $I \otimes G_y$ where
$$
G_y = \pmatrix{1 & 0 \cr 0 & d_y}
$$
and $d_y$ is the diagonal value of $D$ corresponding to state $y$. 
The $\delta_y(x)$ value in the additional register can be removed by 
repeating the transform $U_{\delta_y}$.
This argument shows how a general diagonal matrix can be implemented.
Note that this
implementation is not an efficient one, as it is 
exponential in $n$. 
As we describe in this paper, many special forms of the matrix can be
implemented much more efficiently.
However this implementation will be used in the 
sequel to implement $k\times k$ diagonal
matrices which are part of
efficient implementations discussed later
where $k$ is polynomial in $n$. 

\subsection{A small number of distinct phases}
\label{poly-phases}

This subsection describes a method for efficiently implementing a 
phase change involving only polynomially many distinct phases $r$.
It requires $O(r)$ operations, two calls to $U_f$, and 
$\lceil \log_2(r) \rceil$ additional bits.

Suppose there are $r$ distinct values $p_0,\ldots,p_{r-1}$ of 
$d_x$ such that $r \leq k = 2^m$ for some $k$ that is a power of $2$.  Further
suppose $f(x)$ is a rapidly computable function from $n$-bits to
the values $\{0,\ldots,r-1\}$ such that $d_x = p_{f(x)}$.
Let $P$ be the $k \times k$ diagonal matrix with diagonal elements 
$p_0,\ldots,p_{k-1}$ where $k-r$ elements are chosen arbitrarily. 
Starting with the superposition
$$\ket{0} = \sum_x a_x \ket{x,0},$$
we first apply the transform of \eq{Uf}, with the result 
given in \eq{transform}. We then operate with 
$I \otimes P$ on this result, giving
\begin{equation}
	\sum_x p_{f(x)} a_x \ket{x, f(x)}.
\end{equation}
Finally, the extra register for the index can be disentangled by reversing 
the computation of the index. Since bitwise exclusive-{\sc or} is its own 
inverse, this disentangling can be accomplished by redoing the $U_f$
operation, giving
\begin{equation}
	 \sum_x p_{f(x)} a_x \ket{x, f(x)} \rightarrow \sum_x p_{f(x)} a_x \ket{x, 0},
\end{equation}
which is the desired phase change.

This algorithm requires two evaluations of the $U_f$. In addition
to depending on the efficiency with which $f(x)$ can be computed, this
algorithm depends on the efficiency of implementations for the matrix $P$.
As was shown above, the direct evaluation of 
a $k \times k$ diagonal matrix costs at most $O(k)$.
Note that the cost 
to implement $I \otimes P$ is the same as 
that to implement just $P$. 

Working with the $k \times k$ diagonal matrix $P$ of 
the distinct phase choices, 
instead of the full $N \times N$ diagonal matrix $D$ reduces the cost 
of implementing $D$. In particular,
when $k$ depends polynomially on $n$, the matrix $D$ can be implemented
in polynomial time using $P$, even though the size of the matrix $D$
itself increases exponentially with $n$.

\subsection{Roots of unity}
\label{unity-phases}

When the desired phases are roots of unity, $D$ can be implemented
somewhat more efficiently. 
By using fewer operations than the general case given above, these alternate
techniques will likely be somewhat less sensitive to errors, in addition
to the advantage of faster operation. 

\subsubsection{Changing the sign}
\label{sign-change}

The following technique was introduced by Boyer et al~\cite{Boyer-et-al-96}.  
Let $f(x)=1$ if the sign of $x$ is to change, and $f(x)=0$ otherwise. The
additional register is set to the superposition $\ket{a} =
\frac{1}{\sqrt{2}}(\ket{0} - \ket{1})$.  The operation $U_f$ of
\eq{Uf} then gives a superposition in which the phase of those $x$
with $f(x) = 1$ are inverted and $\ket{a}$ remains unchanged.  This is
readily seen as follows:
\begin{eqnarray*}
\lefteqn{U_f\left(\sum_{x} a_x \ket x \otimes {1\over \sqrt{2}}(\ket 0 - \ket 1)\right) }\\
 &=& {1\over \sqrt{2}} \left(\sum_{x\in X_0} a_x \ket{x,0} - \sum_{x\in X_0} a_x \ket{x,1} + \sum_{x\in X_1} a_x \ket{x,1} - \sum_{x\in X_1} a_x \ket{x,0} \right)\\ 
&=& \left(\sum_{x\in X_0} a_x \ket{x} - \sum_{x\in X_1} a_x \ket{x} \right)\otimes {1\over \sqrt{2}}(\ket{0} - \ket 1)\\
\end{eqnarray*}
where $X_0 = \{x | f(x) = 0\}$ and $X_1 = \{x | f(x) = 1\}$.
The operation introduces a phase
factor of $-1$ for exactly those $x\in X_1$, as desired.  It also leaves
$\ket{a}$ unchanged. In particular the extra register is not entangled
with the $x$ values.

This technique requires only one call to $U_f$, but restricts the phases to 
$1$ and $-1$. Otherwise it requires the same number of resources as the
method described in section \ref{poly-phases}. The method described here
can be generalized somewhat, to phases which are $2^m$th roots of unity, but
it cannot be generalized to arbitrary phase values.

\subsubsection{No direct generalization to arbitrary phase values}

Suppose we want to change the phase of all of the elements of $X_1$ by
$\gamma$. Instead of using 
$\ket{a} = {1\over \sqrt{2}}(\ket 0 - \ket 1)$ we use 
$\ket{a} = {1\over \sqrt{2}}(\ket 0 + \gamma \ket 1)$.  The
result of applying $U_f$ is
\begin{equation}\label{entangled}
{1\over \sqrt{2}}\left( \sum_{x\in X_0} a_x \ket{x,0} + \gamma \sum_{x\in X_0} a_x \ket{x,1} +
                         \sum_{x\in X_1} a_x \ket{x,1} + \gamma \sum_{x\in X_1} a_x
 \ket{x,0}\right).
\end{equation}
In general, the resulting state is not simply a tensor product 
of $x$ and $a$ with some additional
phase shift. Usually, $x$ and $a$ become entangled. 

A possible approach to extracting the desired state from
 this entanglement is to measure the last bit.
The state in \eq{entangled}
becomes either
$$  \sum_{x\in X_0} a_x \ket{x,0} + \gamma \sum_{x\in X_1} a_x \ket{x,0}$$
or
$$ \gamma \sum_{x\in X_0} a_x \ket{x,1} +  \sum_{x\in X_1} a_x \ket{x,1}.$$
If the measurement returns $0$, we have achieved the desired phase shift.  
To get the desired result when the
measured value is $1$, we try multiplying the state by $\gamma$ to
get 
$$ \gamma^2\sum_{x\in X_0} a_x \ket{x,1} + \gamma\sum_{x\in X_1} a_x \ket{x,1}.$$
We get the desired result only when $\gamma^2=1$. 

\subsubsection{Phase changes by a $2^m$th root of unity}

While the preceding calculation shows that general phase changes cannot
be implemented with the technique for changing signs, the behavior when
the last bit is measured does suggest
a way to change the phase of the elements of $X_1$ by a $2^m$th
root of unity.  

For example, this trick can be used to rotate part of the state by $i$
or $-i$.  Let $\gamma = i$.  Perform $U_f$ and measure the last
bit.  If the result is $0$, the state will be
$$\sum_{x\in X_0} a_x \ket{x,0} + i \sum_{x\in X_1} a_x \ket{x,0}$$
and if the result is $1$, the result will be 
$$i \sum_{x\in X_0} a_x \ket{x,1} + \sum_{x\in X_1} a_x \ket{x,1} = 
 i \left(\sum_{x\in X_0} a_x \ket{x,1} - i\sum_{x\in X_1} a_x \ket{x,1}\right).$$
Except for a constant factor, the two states differ only in the phase
of $x\in X_1$ and one can be transformed into the other by applying a
phase change of $-1$ to $X_1$. 
Thus half the time, when $0$ is measured, only one call to $f(x)$
is needed. Otherwise a second phase change is needed, which requires
an additional call to $f(x)$ for a total of two calls.

By iterating this process, one can achieve arbitrary rotations by 
$2^m$th roots of unity.  Let $\gamma = e^{2\pi i/2^m}$. The transformation and measurement of the last bit give
$$\sum_{x\in X_0} a_x \ket{x,0} + e^{2\pi i/2^m} \sum_{x\in X_1} a_x \ket{x,0}$$
or
$$e^{2\pi i/2^m} \sum_{x\in X_0} a_x \ket{x,1} + \sum_{x\in X_1} a_x \ket{x,1}$$
when the last bit is measured to be 0 or 1, respectively.  In the
latter case the state is, up to a constant overall phase, 
$$ \sum_{x\in X_0} a_x \ket{x,1} + e^{-2\pi i/2^m}\sum_{x\in X_1} a_x \ket{x,1}.$$ 
Essentially $X_1$ has been rotated by the right amount, but in the
wrong direction. The desired state can be achieved by rotating 
$X_1$ by $e^{2\pi i/2^{m -1}}$, twice the original amount,
using the same process.  In the worst case, rotating
elements in $X_1$ by $e^{2\pi i/2^m}$ requires $O(m)$ invocations of
$U_f$.  Surprisingly, the average number of calls to $f(x)$ for this
rotation is only $\frac{2^{m-1}-1}{2^{m-2}}$. This average is always
less than two, so on average this technique requires fewer calls than
the method given in section~\ref{poly-phases}.

\subsubsection{$k$th roots of unity}

A different generalization of the sign change technique of
section~\ref{sign-change} allows additional function calls to be
avoided completely.  Furthermore, multiple phases, even up to $2^n$ of
them, can be achieved in this way, as long as they are all multiples of
the same underlying phase $\omega = e^{2\pi i/k}$.  This technique
requires only one function call plus $\log_2(k)$ steps, and $\log_2(k)$
additional qubits.

In this case, the bitwise exclusive-{\sc or} in \eq{Uf} is replaced
by modular addition. Specifically, we use
\begin{equation}\label{Ufmod}
U_f: \ket{x,a} \to \ket{x, a+f(x) \bmod k}.
\end{equation}
Here, $f(x)$ maps states to the set $\{0,\ldots,k-1\}$ and the
desired phase adjustment for state $x$ is $\omega^{f(x)}$, where $\omega = e^{2 \pi i/k}$. To perform this adjustment with a single evaluation of $f(x)$,
we set the extra register in the superposition
\begin{equation}
R = \frac{1}{\sqrt{k}} \sum_{h=0}^{k-1} \omega^{k-h} \ket{h}.
\end{equation} 
The superposition $R$ can be constructed in $\log k$ steps using the technique 
described in section \ref{decomposition}. 

To see the behavior of $U_f$ of \eq{Ufmod} acting on $S \otimes R$, write
\begin{equation}
	S = \sum_{j=0}^{k-1} \sum_{x \in X_j} a_x \ket{x},
\end{equation}
where $X_j$ is the set of states for which $f(x)=j$. Then
\begin{equation}
S \otimes R  =  \frac{1}{\sqrt{k}} 
	\sum_h \sum_j \sum_{x \in X_j} a_x \omega^{k-h} \ket{x, h}.
\end{equation}

Operating with \eq{Ufmod} then gives
\begin{equation}
\frac{1}{\sqrt{k}} 
	\sum_h \sum_j \sum_{x \in X_j} a_x \omega^{k-h} \ket{x, h+j \bmod k}.
\end{equation}
For any $j$, as $h$ ranges from 0 to $k-1$, $m = h+j \bmod k$ ranges over 
these values as well. In terms of $m$, $h = m-j \bmod k$ 
and $k-h = j+(k-m) \bmod k$. Furthermore, since $\omega^k=1$, 
we can write the sum as
\begin{equation}
\frac{1}{\sqrt{k}} 
	\sum_m \sum_j \sum_{x \in X_j} a_x \omega^j \omega^{k-m} \ket{x, m}
\end{equation}
or
\begin{equation}
\frac{1}{\sqrt{k}} 
	\sum_j \sum_{x \in X_j} a_x \omega^j \ket{x} \otimes \sum_m \omega^{k-m} \ket{m},
\end{equation}
which is just $D S \otimes R$.

\section{Approximation of Phase Changes}
\label{approximation}

An arbitrary phase can be approximated by a series of shifts by roots
of unity.  For instance, consider $\phi = e^{p2\pi i}$ for $0 \leq p <
1$.  Let $p = 0.b_1b_2\dots b_k$ be the binary expansion of $p$ to the
desired precision.  Then
\begin{equation}\label{iter}
\phi = \exp \left( 2\pi i \sum_{j=1}^k b_j 2^{-j} \right) = 
	\prod_{j\in B} e^{2\pi i 2^{-j}}
\end{equation}
where $B = \{j | b_j = 1\}$.

Knill \cite{Knill-95} shows that
arbitrary unitary transformations cannot be efficiently approximated.
However, if the phase changes can be concisely described, then they
can be approximated to $k$ bit precision using \eq{iter}. Let 
the phase change be represented by
a diagonal matrix $D$ with phases $D_{mm} = p_m$,
and let $f_j$ for each $j<k$ be such that $f_j(m)$ is the $j$-th 
bit of $p_m$. Then $D$ can be implemented
to $k$ bit precision using one evaluation of each $f_j$.  
This can be done by using one of the techniques described 
in section \ref{independent}
for each $f_j$ using the $2 \times 2$ phase matrices 
$$\left(\begin{array}{cc}1 & 0\\0 & e^{2\pi i/2^j}\end{array}\right).$$
Thus, an arbitrary diagonal matrix $D$ can be approximated 
to $e^{2\pi i/2^k}$ in $O(k)$ steps plus the time it takes to compute
each of the $f_j$'s.


\section{Decomposition}
\label{decomposition}
A diagonal matrix $D$ of size $N=2^n$ representing a phase change of an
$n$ qubit system can be implemented in $O(n)$ steps if 
it is decomposable into single-bit rotations on each of the $n$ bits. 
In this section we give a test for decomposability of a matrix $D$ with
diagonal elements $d_j = D_{jj}$. 
As multiplying the entire state by a constant phase factor has no
physical meaning, we may assume that $d_0 = 1$ without loss of generality.

A diagonal matrix $D$ is single-bit-decomposable if $D = G_{n-1}
\otimes \dots \otimes G_0$ where $G_j$ are single bit phase shift
gates of the form
$$G_k= \left(\begin{array}{cc}1 & 0\\0 & g_k\end{array}\right).$$
Thus, the elements $d_j$ are of the form 
$d_j = \Pi_{k=0}^{n-1}g_k^r$, where $r$ is the value of the $k$-th bit 
of the binary expansion of $j$, if and 
only if $D$ is decomposable.  
Equivalently, given the binary representation $j = b_{n-1} \dots b_1b_0$ then 
\begin{eqnarray}
d_j = d_{b_{n-1} \dots b_1b_0} = g^{b_{n-1}}_{n-1} \dots g^{b_1}_1g^{b_0}_0.\label{form-dj}
\end{eqnarray}
In particular it follows that 
\begin{eqnarray}
g_k = d_{2^k} = D_{2^k2^k}.\label{form-gj}
\end{eqnarray}

An effective way to test whether $D$ is decomposable is to see whether
the $g_k$'s given by \eq{form-gj} satisfy \eq{form-dj}.  For arbitrary
phase changes, this test is exponential in $n$, but in most practical
cases the $d_j$'s will be given by some function in terms of which the
test can be performed efficiently.

For any pair $\{x,x'\}$ with $x > x'$ that differ only in bit $k$ of their
binary representations, it follows from \eq{form-dj} and
\eq{form-gj} that $d_x/d_{x'} = g_k = d_{2^k}$.  This condition is necessary 
for decomposability, so can be used as a way to rule out matrices that are not
decomposable.

\section{Permutations}
\label{permutations}
In this section we will discuss efficient ways of implementing permutations.
These transformations are often used in reordering states, so that
subsequent operations can be efficiently implemented. For example, 
many diagonal matrices are decomposable when the states are ordered in 
some appropriate way. In contrast to more general unitary operations,
permutations take each basis vector to another basis vector, rather
than to a superposition of two or more basis vectors.

Every permutation of the $2^n$ basis vectors of an $n$-bit quantum
register corresponds to a classical computation on this register and
vice versa.  To see this note that a Toffoli gate (T) applied to any
$3$ bits of an arbitrary quantum state is a permutation on the basis
vectors.  Since T is complete for all classical computations, all
classical computations are permutations.  On the other hand, each
permutation can be decomposed into a sequence of swaps each of which
can be realized by a classical computation.

We consider permutations that are described by a function $g(x)$ of
the form $U_g: \ket{x, 0} \to \ket{x, g(x)}$ with the requirement that
both the permutation function, $g(x)$, and its inverse $g^{-1}(x)$
must be computable in polynomial time.  These restrictions, which are
stronger than those of previous sections, prevent the efficient
implementation of permutations like the exchange of the desired state
and the state $\ket{00\dots 0}$ described in section \ref{exchange}.


The algorithm itself is simple. Every state computes its destination

\begin{equation}
\sum_x a_x \ket{x,0} \rightarrow \sum_x a_x \ket{x,g(x)}.
\end{equation}

\noindent
after which the $g(x)$ bits erase the $x$ bits. This last step can be 
accomplished using the exclusive-{\sc or} operation and the function
$g^{-1}(x')$:

\begin{equation}
\sum_x a_x \ket{x,g(x)} \rightarrow \sum_x a_x \ket{x \oplus g^{-1}(g(x)),g(x)}
= \sum_x a_x \ket{0,g(x)}.
\end{equation}

\noindent
If the position of the answer is relevant, the right and left parts of the
register can always be exchanged by swapping individual qubits.

The total computation time that this operation requires is just the time
to compute $g(x)$ plus the time to compute its inverse.

Note that this process turns any classical bijection $g$ of the form
$U_g: \ket{x, 0} \to \ket{x, g(x)}$ into an in-place computation of $g$ of the form
$U_{g'}: \ket x \to \ket{g(x)}$.

\section{Mixing Operations}
\label{mixing}

For effective
quantum algorithms, we also need to be able to efficiently mix amplitudes 
in a superposition so as to increase the chance of a desired reading
being made.
One way to achieve this mixing is to combine an efficiently
implementable diagonal matrix with a decomposable mixing matrix.
For instance, a number of
existing algorithms~\cite{STOC::Grover1996,Hogg-97} make use of 
mixing matrices of the form $WDW$ where $D$ is a diagonal matrix 
and $W$ is the Walsh-Hadamard transform given by 
$$ W_{xy} = \frac{1}{2^{n/2}} (-1)^{|x \cdot y|}. $$
We have described efficient implementations for certain diagonal matrices 
that can be combined with the Walsh-Hadamard transformation or other
mixing matrices to achieve desireable amplitude interference.

Another option for efficiently combining amplitudes, described 
in the remainder of the section, combines
permutations with block-diagonal matrices to perform a different class
of mixing operations. These mixing operations partition the standard
basis for quantum computation into small subsets, and mix amplitudes 
only between components in the same partition.

\subsection{Polynomial size block matrices}
\label{blocks}
An extension to the ideas presented so far is to consider matrices
with a few off-diagonal elements. Specifically, we will talk
about block diagonal matrices made out of equally sized $k \times k$ blocks
$\{B_l\}$,
\begin{equation}
M = \pmatrix{B_0 & \cr
		& B_1 \cr
		& & \ddots \cr
		& & & B_{j-1} \cr
}.
\end{equation}

Many of the techniques used for implementing diagonal matrices can also be
used for block matrices. The techniques are particularly useful when 
all the blocks have the same size $k$, because $k$ must then be a power of two
and the blocks act entirely on the lowest $\log_2(k)$ bits. Multiplying
by $M$ is equivalent to the higher bits choosing a unitary matrix
to apply to the lower bits. This is the equivalent of states
choosing a phase when multiplied by a diagonal matrix.

In this section we will expand the technique discussed in
section \ref{poly-phases}. In the diagonal case, we showed how
an exponentially-sized diagonal matrix, could be implemented using a
polynomial-sized diagonal matrix. The only restriction on the original 
matrix was that the number of different phases had to grow polynomially
with the number of bits. 

For block diagonal case, we will do the same. We start with an 
exponentially-sized matrix $M$, and reduce it to a polynomial one.
Instead of restricting the number of distinct phases, we restrict both
the size of the blocks $k$ and the number of distinct blocks $\alpha < j$
which make up $M$ to be polynomial in $n$.
The large matrix $M$, must also be described by a function $f(x)$ that 
determines the locations of the blocks. If the distinct blocks are labeled
with numbers from $0$ to $\alpha-1$, then $f(x)$ assigns to each state
the number of the block in $M$ that would multiply it.
Of course $f(x)$ must assign the same value for any two states that differ only
by their lowest $\log_2(k)$ bits.

With all the definitions in place we can compute

\begin{equation}
\sum_x a_x \ket{x,0} \rightarrow \sum_x a_x \ket{x,f(x)}.
\end{equation}
All that remains is to multiply this state by a polynomial-sized
block diagonal matrix, which can done as follows.
For each value $c$ in the range of $f(x)$, define $g(y)$, for $y\equiv f(x)$,
to be $1$ if $y=c$ and $0$ otherwise. Then, we multiply the low bits of $x$ 
by the matrix $B_y$ if and only if $g(y)=1$. Knill \cite{Knill-95}
shows that any quantum transformation on $\log_2(k)$ qubits
 can be implemented in at most
$O(k^2\log_2(k))$ operations, so the  total number of operations
needed to perform all of these steps is
$O(\alpha k^2\log_2(k))$.

In the end, the bits containing $f(x)$ must be erased. This can be done
with another call to $U_f$. Hence, this algorithm requires
two calls to $U_f$ plus time $O(\alpha k^2(\log_2(k))$, the time it takes to 
perform each of the $\alpha$ multiplications by the $k \times k$ block 
matrices.

Note that this technique is very similar to the ``quantum direct sum'' algorithm
given by H{\o}yer \cite{Hoyer-97}.  The main difference is that 
H{\o}yer does not
require a polynomial number of different blocks, although he hints that his
method can be speeded up in certain cases along the lines we have described
here.  In return function $f$ becomes $f(x) = x \mod m$ 
where $m$ is the size of each block, and so
$f$ can be computed in-place without additional qubits.

\subsubsection{Combining Permutations and Blocks}

By combining permutations with block matrices, we can form
more general mixing matrices. The idea is to divide the states into
sets of $k$ elements, called $k$-sets, and then mix them according to some 
property of the $k$-set.

The first step reorders the states. We assign to each $k$-set 
a unique number called a group number. We also assign to each state a
number from $0$ to $k-1$, called the member ID, that distinguishes it from
the other states in its $k$-set. Using

\begin{equation}
g(x) = group\_number(x) \cdot k + member\_ID(x),
\end{equation}

\noindent
we can apply the permutation $x \rightarrow g(x)$ which will order the
states with blocks corresponding to the $k$-sets.

The second step involves multiplication by a block diagonal matrix, $M$,
made up of $k\times k$ sized blocks.
The choice of blocks in $M$ given by $f(x)$ will depend only on the group
number of each $k$-set. In this fashion, each $k$-set can be mixed in
different ways depending on its properties.

The final step uses the permutation corresponding to $g^{-1}(x)$ to
send the states back to their original order.

Note that this implementation is efficient only if $k$ is
polynomial in $n$ and if $g$, $g^{-1}$, and $f$ are all
efficiently computable.

\section{Conclusions}
\label{conclusions}
In this paper we have discussed a number of non-classical 
programming techniques for quantum computers.  
Several methods for implementing relative phase changes on
components of an $n$-qubit state were described, as well as the 
trade-offs between these methods in terms of numbers of additional
bits, number of calls to $U_f$, and the number of basic operations needed.
Implementations of permutations and of block diagonal matrices
were also described.
Some of these techniques are more general than those used in
currently known quantum algorithms.  The hope is that they will aid in the
development of future quantum algorithms.

\bibliographystyle{h-elsevier}
\bibliography{qc}

\end{document}